\documentclass[11pt, logo, nonumbering]{preprint}
\usepackage[utf8]{inputenc}
\usepackage{microtype}
\usepackage{graphicx}
\usepackage{amsmath}
\usepackage{booktabs}
\usepackage{xcolor}
\usepackage{multirow}
\usepackage{adjustbox}
\definecolor{mydarkblue}{rgb}{0,0.08,0.45}
\usepackage[colorlinks=true,linkcolor=mydarkblue,citecolor=mydarkblue,filecolor=mydarkblue,urlcolor=mydarkblue]{hyperref}
\usepackage{xspace}
\usepackage{cleveref}

\usepackage[style=numeric-comp,maxbibnames=3,minnames=1,backend=bibtex, doi=false,eprint=false,url=false]{biblatex}

\definecolor{gred}{RGB}{250, 210, 207}
\definecolor{coolblue1}{rgb}{0.91, 0.94, 0.98}
\definecolor{coolblue2}{rgb}{0.76, 0.85, 0.94}
\definecolor{coolblue3}{rgb}{0.54, 0.72, 0.87}
\definecolor{coolblue4}{rgb}{1, 1, 1}

\usepackage[textsize=small]{todonotes}
\newcommand{\promptblock}[2][]{\todo[inline,linecolor=black,backgroundcolor=green!10,bordercolor=black,#1]{#2}}

\newcommand{\method}{AudioCapBench\xspace}
\newcommand{\best}[1]{\textbf{#1}}
\newcommand{\second}[1]{\underline{#1}}

\begin{document}

\title{AudioCapBench: Quick Evaluation on Audio Captioning across Sound, Music, and Speech}

\author{
Jielin Qiu, Jianguo Zhang, Zixiang Chen, Liangwei Yang, Ming Zhu, Juntao Tan,  \\
Haolin Chen, Wenting Zhao, Rithesh Murthy, Roshan Ram, Akshara Prabhakar, \\
Shelby Heinecke, Caiming, Xiong, Silvio Savarese, Huan Wang \\
~~~~\\
\textsuperscript{}Salesforce AI Research \\
~~~~\\
\href{https://github.com/SalesforceAIResearch/AudioCapBench}{
  \raisebox{-0.3\height}{\includegraphics[height=0.8cm]{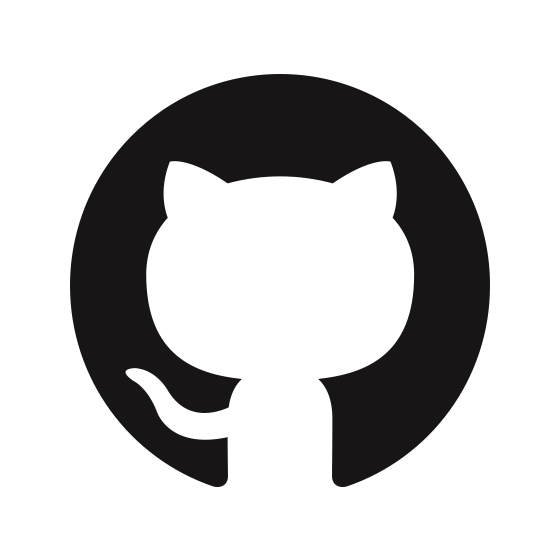}}
  \textbf{https://github.com/SalesforceAIResearch/AudioCapBench}
}
}

\maketitle

\begin{abstract}
We introduce \method, a benchmark for evaluating audio captioning capabilities of large multimodal models. \method covers three distinct audio domains, including environmental sound, music, and speech, with 1,000 curated evaluation samples drawn from established datasets. We evaluate 13 models across two providers (OpenAI, Google Gemini) using both reference-based metrics (METEOR, BLEU, ROUGE-L) and an LLM-as-Judge framework that scores predictions on three orthogonal dimensions: \textit{accuracy} (semantic correctness), \textit{completeness} (coverage of reference content), and \textit{hallucination} (absence of fabricated content). Our results reveal that Gemini models generally outperform OpenAI models on overall captioning quality, with Gemini~3~Pro achieving the highest overall score (6.00/10), while OpenAI models exhibit lower hallucination rates. All models perform best on speech captioning and worst on music captioning. We release the benchmark as well as evaluation code to facilitate reproducible audio understanding research.
\end{abstract}

\section{Introduction}
\label{sec:intro}

Audio captioning, the task of generating natural language descriptions of audio content, is a fundamental capability for multimodal AI systems. Unlike automatic speech recognition (ASR), which transcribes spoken words, audio captioning requires understanding the full acoustic scene: identifying sound sources, describing musical attributes, characterizing speaker emotions, and conveying temporal relationships between audio events.

Recent large multimodal models (LMMs) have introduced native audio input capabilities, enabling direct audio-to-text generation. However, systematic evaluation of these models on audio captioning remains limited. Existing benchmarks~\cite{drossos2020clotho,kim2019audiocaps} focus primarily on training specialized captioning models rather than evaluating general-purpose LMMs, and they typically cover only environmental sounds.

We introduce \method to address these gaps:

\begin{itemize}
    \item \textbf{Multi-domain coverage.} We evaluate across three audio categories: environmental sound (400 samples), music (300 samples), and speech (300 samples), providing a holistic view of audio understanding.
    \item \textbf{Comprehensive model comparison.} We benchmark 13 models across OpenAI (Chat Completions and Realtime models) and Google Gemini (2.0 through 3.x), including both full-size and lightweight variants.
    \item \textbf{Principled evaluation.} We design an LLM-as-Judge framework with three orthogonal dimensions grounded in information retrieval theory (precision, recall, false positive rate), complemented by traditional reference-based metrics.
    \item \textbf{Full reproducibility.} All data, code, and evaluation scripts are publicly released.
\end{itemize}

\section{AudioCapBench}

\subsection{Dataset Construction}

\method comprises 1,000 audio samples across three categories:

\begin{itemize}
    \item \textbf{Sound} (400 samples): 200 from Clotho v2 test split~\cite{drossos2020clotho} and 200 from AudioCaps test split~\cite{kim2019audiocaps}. Environmental sounds including nature, urban, household, and industrial audio.
    \item \textbf{Music} (300 samples): From the MusicCaps evaluation set~\cite{agostinelli2023musiclm}. Professional musician annotations describe genre, instrumentation, tempo, and mood.
    \item \textbf{Speech} (300 samples): From the emotional speech caption dataset~\cite{seastar2024emo}, featuring paralinguistic descriptions of speaker characteristics, emotional tone, and speaking style alongside transcripts.
\end{itemize}

\begin{figure}[t]
    \centering
    \includegraphics[width=0.99\textwidth]{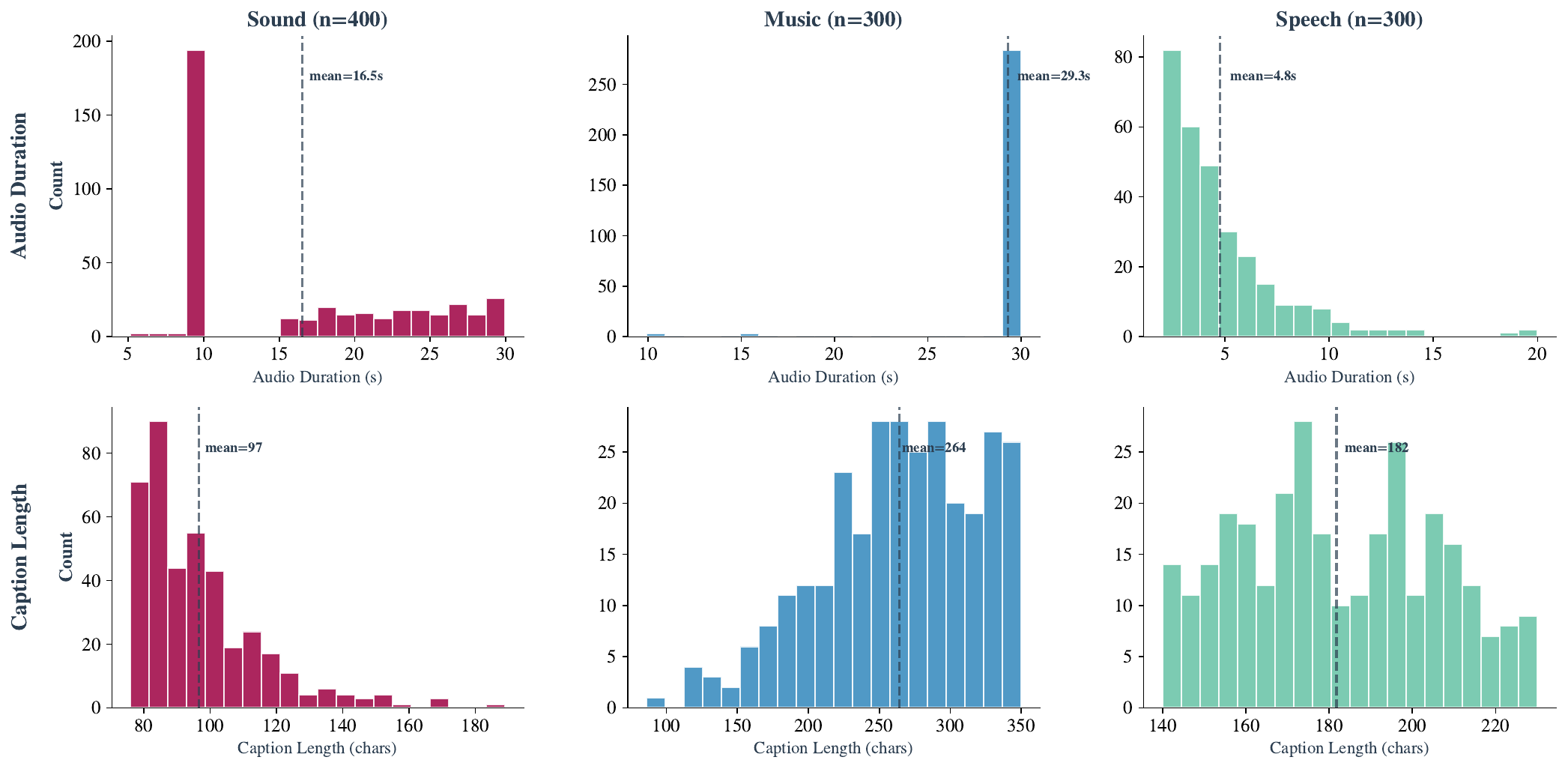}
    \caption{Distribution of audio duration (top) and reference caption length (bottom) for each category. Dashed lines indicate means. Music clips are longest; speech clips are shortest but have detailed emotional captions.}
    \label{fig:dataset}
\end{figure}

Samples were curated for balanced caption lengths within each category (\Cref{fig:dataset}). Evaluation samples were randomly selected from specific caption length ranges using a fixed seed for reproducibility.

\subsection{Evaluation Metrics}

\paragraph{LLM-as-Judge (Primary).}
We use GPT-4.1 as an automated judge, scoring each prediction on three orthogonal dimensions (0--10 scale):

\begin{enumerate}
    \item \textbf{Accuracy} $s_{\text{acc}}$: Does the prediction correctly describe the audio content? Focuses on semantic correctness of described elements, not lexical overlap with references.
    \item \textbf{Completeness} $s_{\text{comp}}$: Does the prediction cover all key elements mentioned in the references? Measures recall of important audio events and attributes.
    \item \textbf{Hallucination} $s_{\text{hall}}$: Does the prediction avoid inventing content not supported by the references? Scored inversely: $10$ = no hallucination, $0$ = heavy hallucination.
\end{enumerate}

For each sample $i$, the judge produces scores $s_{\text{acc}}^{(i)}, s_{\text{comp}}^{(i)}, s_{\text{hall}}^{(i)} \in \{0, 1, \ldots, 10\}$. The per-sample overall score is the simple average:
\begin{equation}
    s_{\text{overall}}^{(i)} = \frac{1}{3}\left(s_{\text{acc}}^{(i)} + s_{\text{comp}}^{(i)} + s_{\text{hall}}^{(i)}\right).
\end{equation}
Corpus-level scores are computed by averaging over all $N$ successfully evaluated samples:
\begin{equation}
    S_d = \frac{1}{N} \sum_{i=1}^{N} s_d^{(i)}, \quad d \in \{\text{acc}, \text{comp}, \text{hall}, \text{overall}\}.
\end{equation}

These three dimensions are designed for orthogonality, each capturing an independent failure mode that maps to a confusion matrix over audio content elements:

\begin{center}
\begin{tabular}{l|cc}
\toprule
 & \textbf{In audio} & \textbf{Not in audio} \\
\midrule
\textbf{Model describes} & Accuracy (precision) & Hallucination (FPR) \\
\textbf{Model omits} & Completeness (recall) & --- \\
\bottomrule
\end{tabular}
\end{center}

We evaluated and rejected several alternative dimensions: \textit{fluency} (all LLMs score 9+/10, providing no discrimination between models), \textit{specificity} (overlaps with accuracy---a more specific correct prediction already scores higher on accuracy), \textit{temporal awareness} (only applicable to ${\sim}50\%$ of samples that contain sequential events), and \textit{descriptive quality} (overlaps with completeness and creates tension with hallucination, as adding detail risks fabrication).

The judge uses category-specific prompts tailored to each audio domain (see \Cref{app:prompt} for full prompts): sound samples are assessed on event identification and acoustic environment; music on genre, instrumentation, tempo, and mood; speech on speaker characteristics, emotion, and transcript content.

\paragraph{Reference-Based Metrics (Secondary).}
We compute three standard metrics against ground-truth reference captions. Given a predicted caption $\hat{c}$ and a set of reference captions $\{r_1, \ldots, r_K\}$ for each sample:

\begin{itemize}
    \item \textbf{METEOR}~\cite{banerjee2005meteor}: Computes unigram alignment between prediction and each reference using exact match, stemming, and synonym matching via WordNet. The score balances precision and recall with a penalty for fragmentation. We report $\max_k \text{METEOR}(\hat{c}, r_k)$.
    \item \textbf{BLEU-4}~\cite{papineni2002bleu}: Measures modified $n$-gram precision (up to 4-grams) of the prediction against all references, with a brevity penalty for short predictions. We apply Method~1 smoothing to handle zero counts.
    \item \textbf{ROUGE-L}~\cite{lin2004rouge}: Computes the longest common subsequence (LCS) F1 between prediction and each reference using stemming. We report $\max_k \text{ROUGE-L}(\hat{c}, r_k)$.
\end{itemize}

These metrics are deterministic, cost-free, and widely used in captioning evaluation, but fundamentally limited by their reliance on surface-level lexical overlap, i.e., a model that says ``fireworks exploding'' when the reference says ``multiple explosions going off'' receives a low score despite being semantically correct.

\subsection{Models Evaluated}

We evaluate 13 models across two providers and three API types:

\begin{itemize}
    \item \textbf{OpenAI Chat Completions} (4 models): \texttt{gpt-4o-audio-preview}, \texttt{gpt-4o-mini-audio-preview}, \texttt{gpt-audio}, \texttt{gpt-audio-mini}
    \item \textbf{OpenAI Realtime} (3 models): \texttt{gpt-4o-realtime-preview}, \texttt{gpt-realtime}, \texttt{gpt-realtime-mini} (WebSocket API, minimum temperature 0.6)
    \item \textbf{Google Gemini} (6 models): \texttt{gemini-2.0-flash}, \texttt{gemini-2.5-flash-lite}, \texttt{gemini-2.5-flash}, \texttt{gemini-2.5-pro}, \texttt{gemini-3-flash-preview}, \texttt{gemini-3-pro-preview}
\end{itemize}

All models use native audio input. Inference uses temperature 0.0 for Chat Completions and Gemini, and 0.6 (minimum allowed) for Realtime models. Gemini 2.5+ thinking models use a thinking budget of 1,024 tokens with max output tokens of 8,192 to prevent truncation.

\section{Results}

\subsection{Overall Leaderboard}

\begin{figure}[htp]
    \centering
    \includegraphics[width=0.95\textwidth]{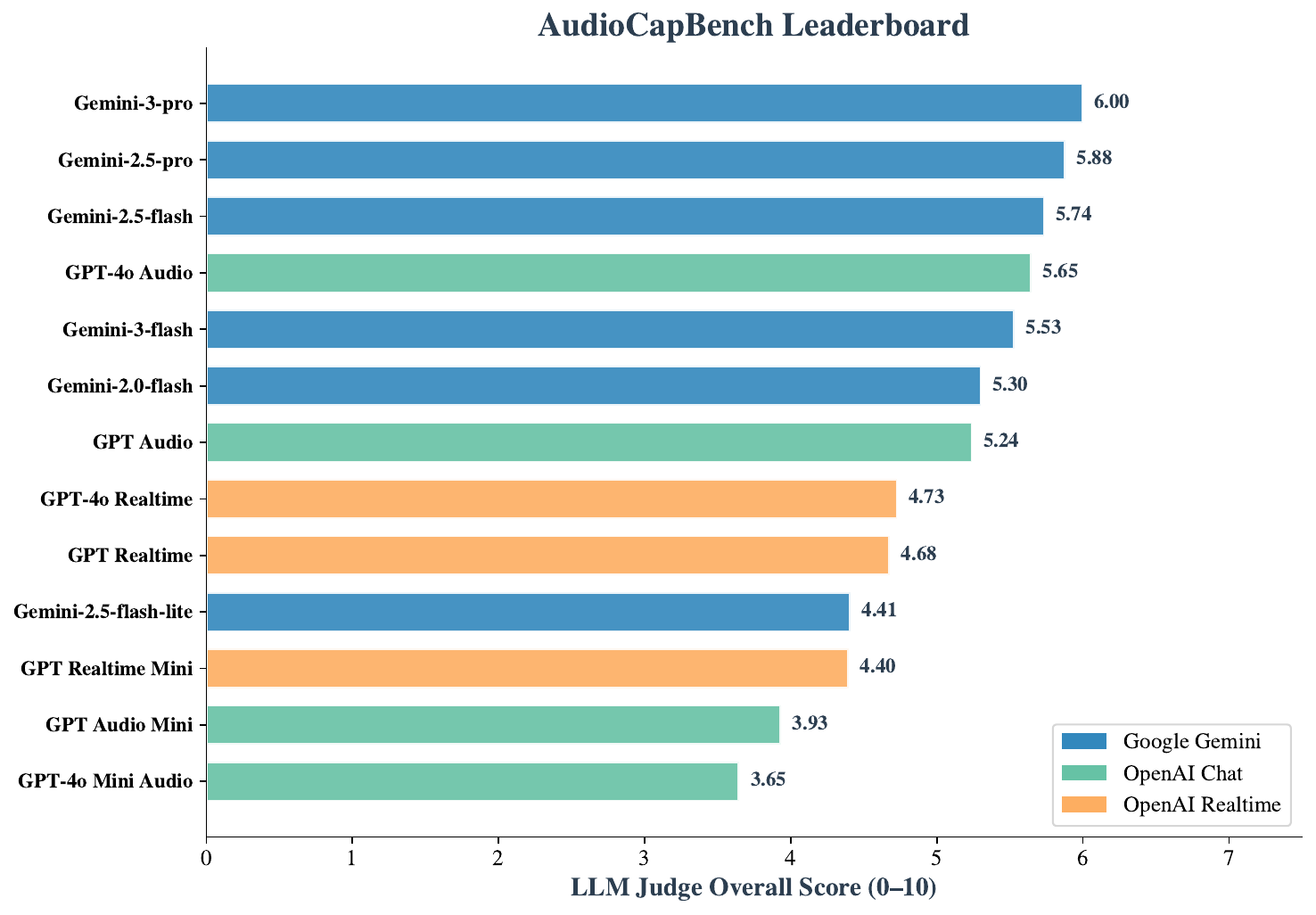}
    \caption{Overall LLM judge scores for all 13 models, colored by provider. Gemini models (blue) dominate the top positions, followed by OpenAI Chat Completions (green) and Realtime (orange).}
    \label{fig:leaderboard}
\end{figure}

\Cref{tab:main} and \Cref{fig:leaderboard} present the full benchmark results. Gemini models occupy the top three positions, with \texttt{gemini-3-pro-preview} achieving the highest overall LLM judge score of 6.00/10.

\begin{table*}[htp]
\centering
\caption{\method results across 13 models on 1,000 samples. LLM judge scores are on a 0--10 scale. \textbf{Bold} = best; \underline{underline} = second best. Sorted by Overall.}
\label{tab:main}
\resizebox{\textwidth}{!}{
\begin{tabular}{clccccccc}
\toprule
\textbf{Rank} & \textbf{Model} & \textbf{Accuracy} & \textbf{Complete.} & \textbf{Halluc.} & \textbf{Overall} & \textbf{METEOR} & \textbf{BLEU-4} & \textbf{ROUGE-L} \\
\midrule
1 & gemini-3-pro-preview & \best{6.18} & \best{6.39} & 5.42 & \best{6.00} & 0.234 & \second{0.037} & 0.152 \\
2 & gemini-2.5-pro & \second{6.00} & \second{6.27} & 5.37 & \second{5.88} & 0.220 & 0.021 & 0.127 \\
3 & gemini-2.5-flash & 5.78 & 5.78 & 5.66 & 5.74 & 0.225 & 0.031 & 0.141 \\
4 & gpt-4o-audio-preview & 5.52 & 5.12 & \second{6.30} & 5.65 & 0.211 & 0.018 & 0.162 \\
5 & gemini-3-flash-preview & 5.68 & 5.81 & 5.11 & 5.53 & \best{0.240} & \best{0.041} & 0.172 \\
6 & gemini-2.0-flash & 5.01 & 4.43 & 6.47 & 5.30 & 0.174 & 0.025 & \best{0.188} \\
7 & gpt-audio & 4.83 & 4.53 & 6.37 & 5.24 & 0.196 & 0.013 & 0.136 \\
8 & gpt-4o-realtime-preview & 4.65 & 4.29 & 5.25 & 4.73 & 0.204 & 0.022 & \second{0.185} \\
9 & gpt-realtime & 4.67 & 4.46 & 4.90 & 4.68 & 0.209 & 0.021 & 0.165 \\
10 & gemini-2.5-flash-lite & 4.28 & 4.03 & 4.91 & 4.41 & 0.184 & 0.014 & 0.124 \\
11 & gpt-realtime-mini & 4.29 & 3.89 & 5.01 & 4.40 & 0.188 & 0.020 & 0.168 \\
12 & gpt-audio-mini & 2.59 & 2.37 & 6.84 & 3.93 & 0.143 & 0.011 & 0.102 \\
13 & gpt-4o-mini-audio-preview & 1.63 & 1.41 & \best{7.90} & 3.65 & 0.107 & 0.010 & 0.103 \\
\bottomrule
\end{tabular}
}
\end{table*}

\subsection{Per-Category Analysis}

\Cref{tab:category} shows the per-category breakdown.

\begin{table}[htp]
\centering
\caption{LLM Overall scores by audio category (0--10). \textbf{Bold} = best per column.}
\label{tab:category}
\begin{tabular}{lccc}
\toprule
\textbf{Model} & \textbf{Sound} & \textbf{Music} & \textbf{Speech} \\
\midrule
gemini-3-pro-preview & \best{6.40} & \best{5.68} & 5.79 \\
gemini-2.5-pro & 5.76 & 5.66 & 6.26 \\
gemini-2.5-flash & 5.47 & 5.19 & \best{6.63} \\
gpt-4o-audio-preview & 5.40 & 5.15 & 6.47 \\
gemini-3-flash-preview & 5.72 & 4.88 & 5.92 \\
gemini-2.0-flash & 4.55 & 5.27 & 6.34 \\
gpt-audio & 5.61 & 5.19 & 4.79 \\
gpt-4o-realtime-preview & 4.91 & 3.53 & 5.69 \\
gpt-realtime & 4.94 & 3.35 & 5.66 \\
gemini-2.5-flash-lite & 4.04 & 3.82 & 5.50 \\
gpt-realtime-mini & 4.55 & 3.41 & 5.17 \\
gpt-audio-mini & 3.65 & 3.39 & 4.85 \\
gpt-4o-mini-audio-preview & 3.90 & 3.37 & 3.58 \\
\bottomrule
\end{tabular}
\end{table}

\begin{itemize}
    \item \textbf{Speech is easiest} for most models, likely because speech content can be partially inferred from the spoken words. Gemini~2.5~Flash achieves the highest speech score (6.63).
    \item \textbf{Music is hardest}, requiring identification of genre, instrumentation, tempo, and mood---attributes that demand specialized musical knowledge. Gemini~3~Pro leads at 5.68.
    \item \textbf{Sound performance varies widely}, from 6.40 (Gemini~3~Pro) to 3.65 (GPT-audio-mini), suggesting environmental sound event detection is a key differentiator among models.
\end{itemize}

\begin{figure}[htp]
    \centering
    \includegraphics[width=0.99\textwidth]{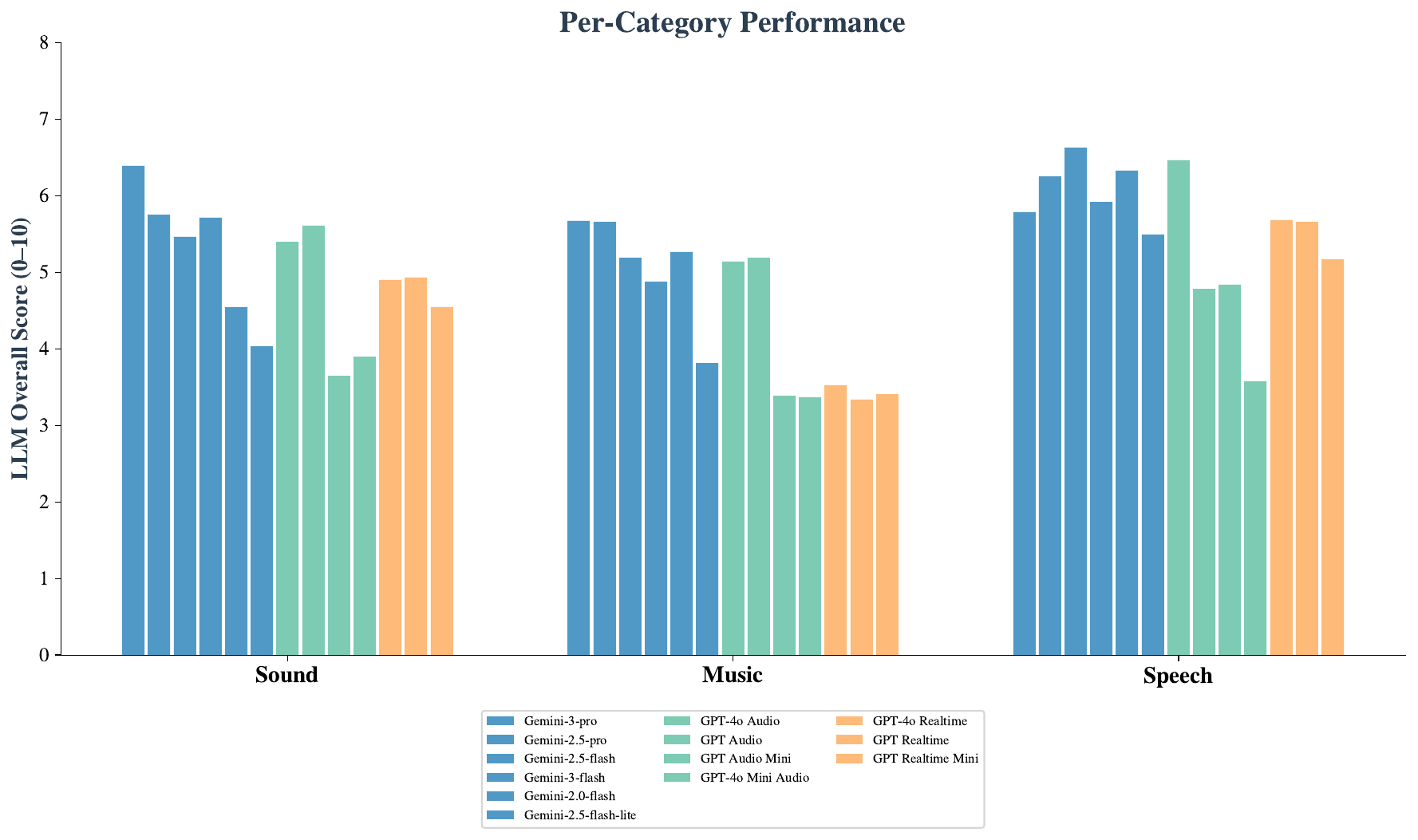}
    \caption{Per-category LLM Overall scores. Speech is consistently the easiest category across all models, while music is the hardest. Performance gaps between models are largest on sound captioning.}
    \label{fig:category}
\end{figure}

\subsection{Findings}

\paragraph{Accuracy--Hallucination Trade-off.}
OpenAI mini models exhibit a striking pattern: for example, \texttt{gpt-4o-mini-audio-preview} achieves the highest hallucination score (7.90, meaning least hallucination) but the lowest accuracy (1.63) and completeness (1.41). These models minimize false positives by producing brief, generic descriptions, sacrificing content coverage entirely. This suggests a conservative design where the model avoids describing uncertain sounds rather than risking incorrect descriptions.

\paragraph{Realtime Models Underperform.}
OpenAI Realtime models consistently score lower than Chat Completions counterparts. This is partly due to the minimum temperature of 0.6 (vs.\ 0.0 for Chat Completions), which introduces variability, and the conversational design of the Realtime API optimized for interactive dialogue rather than precise captioning.

\paragraph{Gemini's Completeness Advantage.}
Gemini models consistently outperform OpenAI on completeness, producing more detailed descriptions. Gemini~3~Pro achieves 6.39 vs.\ 5.12 for GPT-4o-audio-preview. However, this comes at the cost of higher hallucination (5.42 vs.\ 6.30), as more detailed descriptions increase the risk of fabrication.

\paragraph{Reference Metrics vs.\ LLM Judge.}
Traditional metrics show weak correlation with LLM judge scores. Gemini~3~Flash achieves the highest METEOR (0.240) and BLEU-4 (0.041) but ranks only 5th overall. This confirms that lexical overlap metrics are insufficient for audio captioning, where models may correctly describe audio using entirely different vocabulary.

\begin{figure}[htb]
    \centering
    \includegraphics[width=0.99\textwidth]{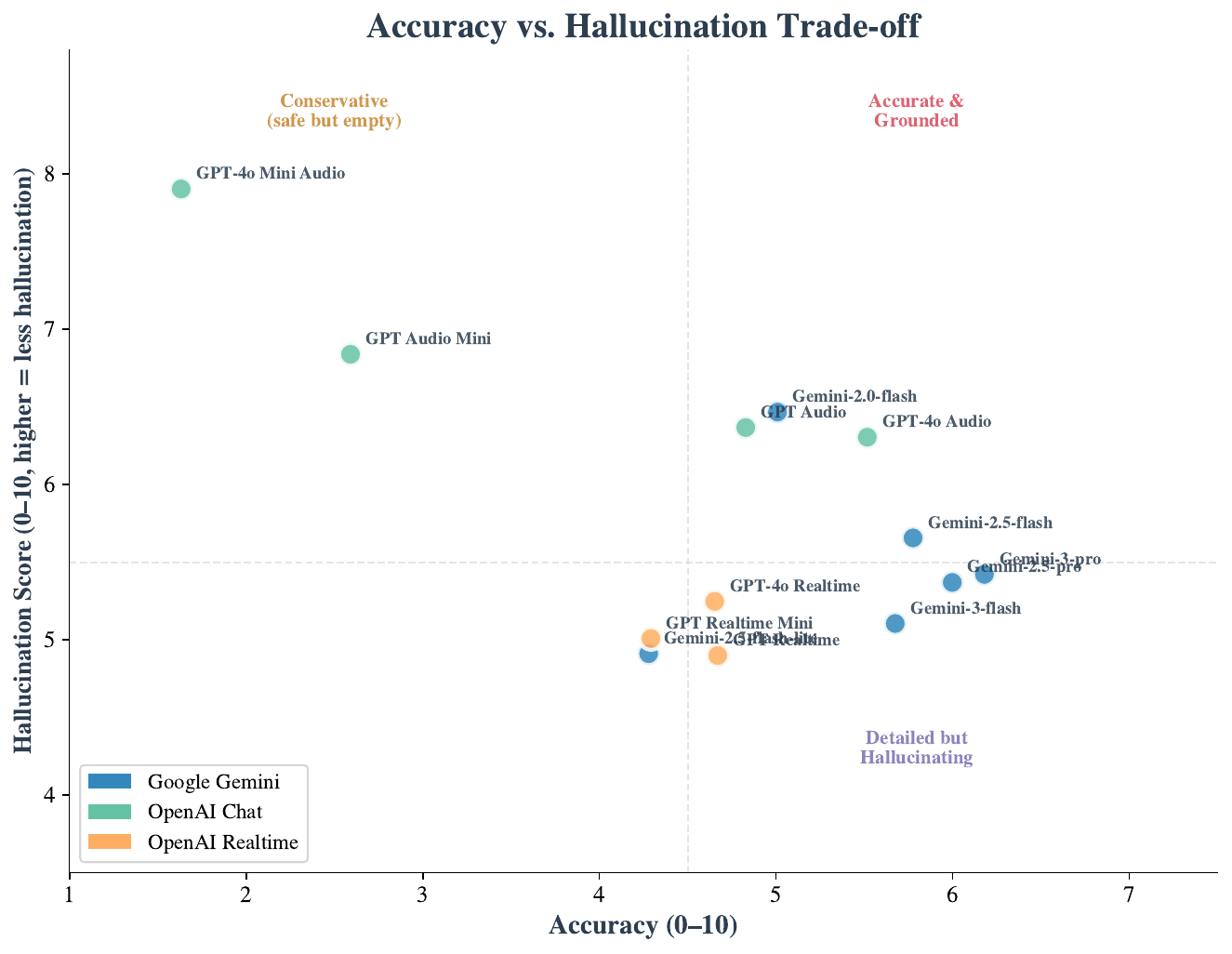}
    \caption{Accuracy vs.\ hallucination trade-off. Models in the upper-right are both accurate and grounded. OpenAI mini models cluster in the upper-left (conservative but empty). Gemini models tend toward higher accuracy but lower hallucination scores.}
    \label{fig:tradeoff}
\end{figure}

\section{Related Work}
\label{sec:related}

\paragraph{Audio Captioning Datasets.}
Clotho~\cite{drossos2020clotho} provides 4,981 audio clips with 5 crowdsourced captions each, drawn from Freesound. AudioCaps~\cite{kim2019audiocaps} extends AudioSet~\cite{gemmeke2017audioset} with human-written captions for 46K audio clips from YouTube. MusicCaps~\cite{agostinelli2023musiclm} offers 5,521 music clips annotated by professional musicians with detailed descriptions of genre, instrumentation, and mood. WavCaps~\cite{mei2024wavcaps} scales up audio-language data through ChatGPT-assisted weak labeling, collecting 403K audio clips with generated descriptions. Auto-ACD~\cite{sun2023autoacd} proposes an automated pipeline to construct audio captioning datasets from AudioSet by generating captions via LLMs conditioned on audio event labels. These datasets were originally designed for training specialized models; we repurpose established test/evaluation splits for benchmarking general-purpose LMMs.

\paragraph{Audio Captioning Models.}
Traditional audio captioning follows an encoder-decoder paradigm, where audio spectrograms are encoded by pretrained audio models (e.g., PANNs~\cite{kong2020panns}, AST~\cite{gong2021ast}) and decoded into captions by recurrent or transformer networks. The DCASE Audio Captioning Challenge~\cite{dcase2023task6a} has driven progress in this area since 2020. More recently, large audio-language models have emerged: SALMONN~\cite{tang2024salmonn} combines a whisper encoder with a BEATs audio encoder and routes both into a Vicuna LLM; Qwen-Audio~\cite{chu2023qwenaudio} trains a unified audio encoder jointly with Qwen-7B across diverse audio tasks; and Pengi~\cite{deshmukh2023pengi} frames all audio tasks as text generation conditioned on audio and text inputs. However, the latest generation of commercial LMMs (GPT-4o-audio, Gemini 2.5/3) accept raw audio natively, bypassing the need for separate audio encoders entirely.

\paragraph{Audio Captioning Evaluation.}
Standard evaluation in audio captioning relies on reference-based metrics borrowed from machine translation and image captioning. SPIDEr~\cite{liu2017spider}, the average of SPICE and CIDEr-D, has been the primary ranking metric in the DCASE challenges. FENSE~\cite{zhou2022fense} augments sentence-BERT similarity with a fluency error detector and was adopted as the DCASE 2024 ranking metric. However, all reference-based metrics share a fundamental limitation: they penalize semantically correct but lexically different predictions.

\paragraph{LLM-as-Judge Evaluation.}
Using LLMs to evaluate generated text has gained popularity as an alternative to reference-based metrics. Zheng et al.~\cite{zheng2023judging} introduce MT-Bench and demonstrate strong agreement between GPT-4 judges and human annotators on open-ended generation. Subsequent work has explored position bias~\cite{wang2023large}, calibration~\cite{ye2024justice}, and multi-judge protocols. For audio specifically, AudioBench~\cite{wang2024audiobench} uses GPT-4 to evaluate audio understanding across speech, sound, and music tasks but focuses on question-answering rather than free-form captioning. Our LLM-as-Judge framework is specifically designed for audio captioning with three orthogonal dimensions grounded in information retrieval theory.

\section{Conclusion}
\label{sec:conclusion}

We presented \method, a benchmark for evaluating audio captioning across sound, music, and speech domains. Our evaluation of 13 models reveals: (1) Gemini models lead overall, with Gemini~3~Pro achieving the top score; (2) all models struggle most with music captioning; (3) a fundamental accuracy--hallucination trade-off exists where conservative models avoid errors but miss content; and (4) traditional metrics poorly reflect caption quality compared to LLM-based evaluation.

\paragraph{Limitations.} We use a single LLM judge (GPT-4.1), which may introduce systematic bias. Reference captions may not capture all valid descriptions of an audio clip. The benchmark currently only evaluated on API-based models and does not include open-weight audio models.

\paragraph{Future Work.} Multi-judge evaluation, reference-free metrics, expansion to additional audio domains, and inclusion of open-weight models.

\paragraph{Acknowledgments.} We thank Jim Jagielski and Jeanette Berberich for their help with the open-source release process.

\clearpage
\printbibliography

\newpage
\appendix

\section{Evaluation Prompts}
\label{app:prompt}

\subsection{LLM Judge Base Prompt}

The following prompt is sent to GPT-4.1 for each sample. The \texttt{\{category\_guidance\}} placeholder is replaced with one of the category-specific prompts in \Cref{app:catprompts}, and \texttt{\{refs\_text\}} and \texttt{\{prediction\}} are filled with the actual reference captions and model prediction.

\promptblock[backgroundcolor=blue!5,bordercolor=blue!40,linecolor=blue!40]{%
\small\ttfamily
You are an expert evaluator for audio captioning systems.\\[4pt]
Given the ground-truth reference captions and a model's predicted caption for an audio clip, score the prediction on the following criteria (each on a scale of 0 to 10):\\[4pt]
1. \textbf{Accuracy} (0-10): Does the prediction correctly describe the same audio content as the references? Are the key sound sources, events, or attributes correct? Note: the prediction may use different wording than the references --- focus on whether the semantic content is correct, not exact word matches.\\[2pt]
2. \textbf{Completeness} (0-10): Does the prediction cover the main elements mentioned in the references? Are important details missing? A prediction that captures the most salient elements should score highly even if it misses minor details.\\[2pt]
3. \textbf{Hallucination} (0-10): Does the prediction ONLY describe sounds/events that are actually supported by the references? 10 = no hallucination (everything described matches the references), 0 = heavy hallucination (the prediction invents sounds, events, or attributes not present in the references). Penalize any fabricated content, even if the prediction also contains correct elements.\\[4pt]
\{category\_guidance\}\\[4pt]
Reference captions (ground-truth):\\
\{refs\_text\}\\[4pt]
Model prediction: ``\{prediction\}''\\[4pt]
Important: If the prediction is empty, all scores should be 0. Scores must be integers from 0 to 10. Focus on semantic similarity, not surface-level word overlap.\\[4pt]
Respond with ONLY a JSON object, no other text:\\
\{``accuracy'': <int 0-10>, ``completeness'': <int 0-10>, ``hallucination'': <int 0-10>, ``reasoning'': ``<1-2 sentence explanation>''\}
}

\subsection{Category-Specific Guidance}
\label{app:catprompts}

The following guidance is inserted into the base prompt depending on the audio category:

\vspace{-5pt}
\paragraph{Sound (Environmental).}
\promptblock[backgroundcolor=orange!8,bordercolor=orange!50,linecolor=orange!50]{%
For general/environmental sound, evaluate whether the prediction correctly identifies: sound sources, events, acoustic environment, and temporal patterns.}

\vspace{-5pt}
\paragraph{Music.}
\promptblock[backgroundcolor=orange!8,bordercolor=orange!50,linecolor=orange!50]{%
For music, evaluate whether the prediction correctly identifies: genre, instrumentation, tempo, mood/atmosphere, and any vocal characteristics.}

\vspace{-5pt}
\paragraph{Speech.}
\promptblock[backgroundcolor=orange!8,bordercolor=orange!50,linecolor=orange!50]{%
For speech, evaluate whether the prediction correctly describes: speaker characteristics (gender, age, accent), emotional tone, speaking style, and the content of what is being said. The reference may include both a transcript and an emotional description---evaluate how well the prediction captures both aspects.}

For speech samples, the actual transcript is additionally appended: \textit{Actual transcript of the speech: ``\{transcript\}''}

\subsection{Captioning Instruction Prompts}

Each model receives one of the following category-specific instructions (alternated per sample):

\vspace{-5pt}
\paragraph{Sound.}
\promptblock[backgroundcolor=green!8,bordercolor=green!50,linecolor=green!50]{%
(1) Describe what you hear in this audio.\\
(2) What sounds can you identify in this audio clip?}

\vspace{-5pt}
\paragraph{Music.}
\promptblock[backgroundcolor=green!8,bordercolor=green!50,linecolor=green!50]{%
(1) Describe this music clip in detail, including genre, instrumentation, tempo, and mood.\\
(2) Characterize this musical excerpt with rich detail; cover genre, instrumentation, and overall atmosphere.}

\vspace{-5pt}
\paragraph{Speech.}
\promptblock[backgroundcolor=green!8,bordercolor=green!50,linecolor=green!50]{%
(1) Describe the speaker and what they are saying, including their tone, emotion, and speaking style.\\
(2) Describe this speech audio, including the speaker's characteristics and what is being said.}

\section{Dataset Statistics}
\label{app:data}

\begin{table}[h]
\centering
\caption{Evaluation dataset composition.}
\begin{tabular}{lcccc}
\toprule
\textbf{Category} & \textbf{Source} & \textbf{Samples} & \textbf{Avg Caption} \\
\midrule
Sound & Clotho v2 test & 200 & 90 chars \\
Sound & AudioCaps test & 200 & 104 chars \\
Music & MusicCaps eval & 300 & 264 chars \\
Speech & Emo Speech Caption & 300 & 182 chars \\
\midrule
\textbf{Total} & & \textbf{1,000} & \\
\bottomrule
\end{tabular}
\end{table}

\end{document}